\documentclass[sigconf]{acmart}

\usepackage{multirow}
\usepackage{soul, color}
\usepackage{subfigure}
\usepackage{booktabs} 
\usepackage[ruled,vlined,linesnumbered]{algorithm2e}
\usepackage{url}
\usepackage{algorithmic}
\usepackage{booktabs} 
\usepackage{xspace}
\usepackage{enumitem}
\usepackage{array}
\usepackage{balance}

\definecolor{selectiveyellow}{rgb}{1.0, 0.73, 0.0}
\definecolor{bluefrance}{rgb}{0.19, 0.55, 0.91}
\definecolor{ao}{rgb}{0.55, 0.71, 0.0}

\setcopyright{rightsretained}

\newcommand{\eat}[1]{}

\newcommand{\ie}{\emph{i.e.,}\xspace}
\newcommand{\eg}{\emph{e.g.,}\xspace}

\newcommand{\baby}{\textsc{CARP}\xspace}
\newcommand{\paratitle}[1]{\noindent\textbf{#1}}

\newcommand{\hlgol}[1]{{\sethlcolor{selectiveyellow}\hl{#1}}}
\newcommand{\hlbl}[1]{{\sethlcolor{bluefrance}\hl{#1}}}

\setul{}{1pt}
\newcommand{\ulred}[1]{{\setulcolor{red}\ul{#1}}}
\newcommand{\ulgrn}[1]{{\setulcolor{ao}\ul{#1}}}

\begin{document}
\copyrightyear{2019} 
\acmYear{2019} 
\setcopyright{acmcopyright}
\acmConference[SIGIR '19]{Proceedings of the 42nd International ACM SIGIR Conference on Research and Development in Information Retrieval}{July 21--25, 2019}{Paris, France}
\acmBooktitle{Proceedings of the 42nd International ACM SIGIR Conference on Research and Development in Information Retrieval (SIGIR '19), July 21--25, 2019, Paris, France}
\acmPrice{15.00}
\acmDOI{10.1145/3331184.3331216}
\acmISBN{978-1-4503-6172-9/19/07}

\fancyhead{}

\settopmatter{printacmref=false, printfolios=false}

\title{A Capsule Network for Recommendation and Explaining What You Like and Dislike}
\author{Chenliang Li$^{1}$, Cong Quan$^2$, Li Peng$^3$, Yunwei Qi$^3$, Yuming Deng$^3$, Libing Wu$^2$}
\affiliation{%
  \institution{
  1. Key Laboratory of Aerospace Information Security and Trusted Computing, Ministry of Education, School of Cyber Science and Engineering, Wuhan University, China \\cllee@whu.edu.cn\\
  2. School of Computer Science, Wuhan University, Wuhan, 430072, China\\\{quancong,wu\}@whu.edu.cn\\ 
  3. Alibaba Group, Hangzhou, China\\\{muchen.pl,yunwei.qyw,yuming.dym\}@alibaba-inc.com}
}

\begin{abstract}
User reviews contain rich semantics towards the preference of users to features of items. Recently, many deep learning based solutions have been proposed by exploiting reviews for recommendation. The attention mechanism is mainly adopted in these works to identify words or aspects that are important for rating prediction. However, it is still hard to understand whether a user likes or dislikes an aspect of an item according to what viewpoint the user holds and to what extent, without examining the review details. Here, we consider a pair of \textit{a viewpoint held by a user} and \textit{an aspect of an item} as a \textit{\textbf{logic unit}}. Reasoning a rating behavior by discovering the informative logic units from the reviews and resolving their corresponding sentiments could enable a better rating prediction with explanation. 

To this end, in this paper, we propose a \textbf{ca}psule network based model for \textbf{r}ating \textbf{p}rediction with user reviews, named \baby. For each user-item pair, \baby is devised to extract the informative logic units from the reviews and infer their corresponding sentiments. The model firstly extracts the viewpoints and aspects from the user and item review documents respectively. Then we derive the representation of each logic unit based on its constituent viewpoint and aspect. A sentiment capsule architecture with a novel \textit{Routing by Bi-Agreement} mechanism is proposed to identify the informative logic unit and the sentiment based representations in user-item level for rating prediction. Extensive experiments are conducted over seven real-world datasets with diverse characteristics. Our results demonstrate that the proposed \baby obtains substantial performance gain over recently proposed state-of-the-art models in terms of prediction accuracy. Further analysis shows that our model can successfully discover the interpretable reasons at a finer level of granularity.
\end{abstract}

%
%
\eat{

}


\maketitle

{\fontsize{8pt}{8pt} \selectfont
\textbf{ACM Reference Format:}\\
Chenliang Li, Cong Quan, Li Peng, Yunwei Qi, Yuming Deng, Libing Wu. 2019. A Capsule Network for Recommendation and Explaining What You Like and Dislike. In {\it Proceedings of the 42nd Int'l ACM SIGIR Conference on Research and Development in Information Retrieval (SIGIR'19), July 21--25, 2019, Paris, France.} ACM, NY, NY, USA, 10 pages. \url{https://doi.org/10.1145/3331184.3331216}
}

\section{Introduction}\label{sec:intro}
\begin{figure}
\centering
\includegraphics[height=30mm,width=0.98\linewidth]{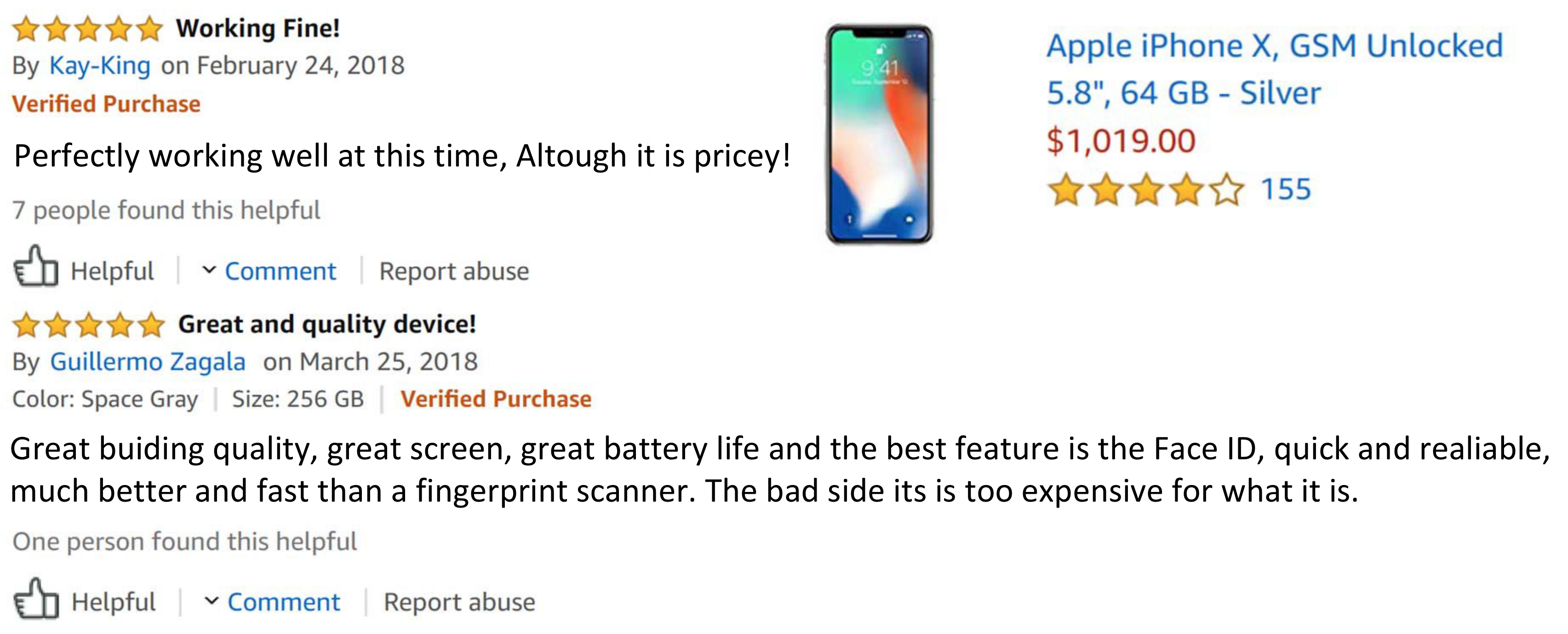}
\caption{Review examples for \textit{Apple iPhone X} in Amazon.}\label{fig:intro}
\end{figure}

Many E-Commerce platforms allow a user to share her experience of a purchased item in the form of review, along with a numerical rating, such as Yelp and Amazon (ref. Figure~\ref{fig:intro}). These textual reviews  have played an increasingly important role for E-Commerce platforms in supporting personalized recommendation. Many recommender systems have been developed by exploiting the semantic information covered in reviews~\cite{recsys13:hfht,recsys14:rmr,kdd11:ctr,aaai14:topicmf,ijcai16:rblt,recsys16:convMf,ijcai16:cmle,wsdm17:deepconn,recsys17:transnet,recsys17:dattn,www18:alfm,www18:narre,kdd18:mpcn,ijcai18:cheng} against using naturally sparse user-item ratings, leading to signficant performance gain.

The earlier solutions choose to adopt topic modeling or non-negative matrix factorization over reviews and represent a user/item semantically~\cite{recsys13:hfht,recsys14:rmr,kdd11:ctr,aaai14:topicmf,ijcai16:rblt}. This modeling paradigm is recently suppressed by the revival of deep learning techniques. Specifically, empowered by continuous real-valued vector representations and semantic composition over contextual information, uptodate review-based deep learning models significantly push the frontier of the state-of-the-art further. Examples include DeepCoNN~\cite{wsdm17:deepconn}, D-Attn~\cite{recsys17:dattn}, TransNet~\cite{recsys17:transnet}, ANR~\cite{cikm18:anr} and MPCN~\cite{kdd18:mpcn}.

Though significant performance gain is achieved by these efforts, these models do not explain what really happens in a user's mind when making a rating. It is reasonable to assume that a user would attach different importance to different aspects of an item. Similarly, different users would care a specific aspect of an item to different extent~\cite{www18:alfm}. When we further consider the sentiment information, the things become more interesting. A user could hold opposite opinions towards an item by considering different aspects.
Hence, the whole rating score can be considered as an indication of an item's relative merit (\ie a compromise on the virtues and achilles heels). The examples demonstrated in Figure~\ref{fig:intro} fall in this line. Here, we refer to all reviews wirtten by a user as \textit{user document}. A user document mainly contains her personal viewpoints towards different items. For example, a user could prefer \textit{outdoor activities} and therefore appreciate the related aspects (\eg \textit{low profile, easy storage}) for an item. Also, an \textit{item document} can be formed by merging all reviews written for the item. An item document can be considered as a summary of the item's various aspects, where the important aspects can be easily exposed~\cite{ijcai16:rblt,www18:tarmf}. Here, to reason a rating behavior, we wish to understand \textit{whether a user likes or dislikes an aspect of an item according to what viewpoint she holds, and to what extent}.

To this end, in this paper, we propose a \textbf{ca}psule network based model for \textbf{r}ating \textbf{p}rediction with user reviews, named \baby. In detail, \baby consists of two components: \textit{viewpoint and aspect extraction} and \textit{sentiment capsules}. Firstly, a variant of self-attention stacked over a convolutional layer is adopted to extract the viewpoints and aspects mentioned from the user and item documents respectively. Given a user-item pair, we pack a user viewpoint and an item aspect together to form a \textit{logic unit}. This notion serves as a proxy for causes behind a rating behavior. However, not all logic units formed by this simple pairing would make sense in real-world. We refer to the semantically plausible logic units as \textit{informative logic units}. For example, a viewpoint about \textit{outdoor enthusiast} and an item aspect about \textit{easy storage} would form an informative logic unit. On the other hand, the pair of the viewpoint about \textit{elegant appearance} and the same aspect about \textit{easy storage} would just result in a horrible haze.

To identify the informative logic unit, we derive the representations of all possible logic units and feed them into the sentiment capsules. Specifically, we utilize a positive capsule and a negative capsule to represent the user's attitudes towards the item. By devising a new iterative routing mechanism - \textit{Routing by Bi-Agreement} - we enable the two sentiment capsules to jointly identify the informative logic units with positive and negative sentiments held by the user respectively. The output vectors produced by these two sentiment capsules encode to what extent the user likes or dislikes the item respectively. Meanwhile, the vector lengths suggest the probability of each of these two sentiments. For a given user-item pair, the overall rating is then estimated based on these magnitudes and odds in two sentiment poles. At last, we need to emphasize that \baby works in a review-driven and end-to-end fashion. To guarantee the proper sentiment modeling, we cast \baby as a multi-task learning process such that implicit sentiment information provided by a rating score is exploited. Specifically, we introduce a sentiment matching process by classifying the sentiment of a review according to its rating score. Note that no human labeling or external NLP tool is required to aid the training of \baby.

Our work differs fundamentally from the prior works, since we aim to make a bridge connecting the reviews and rating scores with causes and effects, in terms of user viewpoints, item aspects and sentiments. Back to existing deep learning based solutions, only important words or aspects in the reviews are highlighted by the representation learning process~\cite{recsys17:dattn,www18:tarmf,cikm18:anr,kdd18:mpcn}. It is difficult to infer what rules a user applies in her mind and the effects separately. Overall, our key contributions are summarized as below:

\begin{itemize}
\item[$\bullet$] We propose a novel deep learning model that exploits reviews for rating prediction and explanation. This is the very first attempt to explicitly model the reasoning process underlying a rating behavior in E-Commerce, by jointly considering user viewpoints, item aspects and sentiments. 

\item[$\bullet$] We introduce a capsule-based architecture to jointly identify informative logic units and estimate their effects in terms of sentiment analysis. We further propose Routing by Bi-Agreement, a new iterative dynamic routing process for capsule networks. 

\item[$\bullet$] To enable sentiment modeling without using any external resource, a multi-task learning process is proposed for model optimization, which can successfully exploit the implicit sentiment information provided by a rating score. 

\item[$\bullet$] On seven real-world datasets with diverse characteristics, our results demonstrate the superiority of the proposed \baby in terms of rating prediction accuracy, performance robustness and explanation at fine-grained level.
\end{itemize}

\section{Related Work}\label{sec:related}
In this section, we briefly review three different areas which are highly relevant to our work.

\subsection{Review-based Recommender Systems}

Recent years, capitalizing user reviews to enhance the precision and the interpretability of recommendation have been investigated and verified by many works~\cite{kdd11:ctr,recsys13:hfht,recsys14:rmr,kdd14:jmars,aaai14:topicmf,ijcai16:rblt,recsys16:convMf,ijcai16:cmle,wsdm17:deepconn,recsys17:transnet,recsys17:dattn,www18:alfm,www18:narre,kdd18:mpcn}. In earlier days, many efforts are made to extract semantic features from reviews with the topic modeling techniques~\cite{sigir99:plsa,jmlr03:lda}. These works integrate the latent semantic topics into the factor learning models~\cite{kdd11:ctr,recsys13:hfht,recsys14:rmr,aaai14:topicmf,ijcai16:rblt}. TLFM proposes two separate factor learning models to exploit both sentiment-consistency and text-consistency of users and items~\cite{ijcai17:tlfm}. Then they unify these two views together to conduct rating prediction. However, the proposed solution requires the review of the target user-item pair as input, which is unrealistic and thus turns the recommendation task into a sentiment analysis task. CDL~\cite{kdd15:cdl} proposes to couple SADE over reviews and PMF~\cite{nips07:pmf}. Since the bag-of-word representation schema is used in these works, signficant information loss is expected due to the missing of contextual information. 

Recently, many works are proposed to model contextual information from reviews for better recommendation with deep learning techniques~\cite{wsdm17:deepconn,recsys17:transnet,recsys17:dattn,www18:narre,kdd18:mpcn}. Convolutional neural networks (CNN)~\cite{emnlp14:kim} and Recurrent Neural Network (RNN)~\cite{cs90:elman,interspeech10:mikolov} are mainly used to composite the semantic contextual information into the continuous real-valued vector representation. ConvMF shares the same architecture with CDL and uses CNN to extract item characteristics from the item description~\cite{recsys16:convMf}. TARMF uses the features extracted from both user document and item document to calibrate the latent factors in the factor learning paradigm~\cite{www18:tarmf}. DeepCoNN uses parallel CNN networks to uncover semantic features from user and item documents~\cite{wsdm17:deepconn}. The extracted features are then fed into the Factorization Machine for rating prediction. TransNet augments DeepCoNN with an additional transform layer to infer the representation of the target user-item review that is unavailable during the rating prediction~\cite{recsys17:transnet}. D-Attn leverages global and local attention mechanisms to identify important words in the review documents for rating prediction~\cite{recsys17:dattn}. Note that each review contains different semantic information. NARRE argues that reviews should be treated differently upon different user-item pairs and proposes to employ an attention mechanism to select useful reviews for rating prediction~\cite{www18:narre}. Likewise, MPCN exploits a pointer-based co-attention schema to enable a multi-hierarchical information selection. Both important reviews and their important words can be identified towards better rating prediction in MPCN~\cite{kdd18:mpcn}.

\subsection{Aspect-based Recommender Systems}
Many efforts are devoted to model user opinions from reviews for transparent recommendation, namely aspect-based recommender systems, which can be further divided into two categories. The first category resorts to using the external NLP tools to extract aspects and sentiments from reviews. For example, EFM and MTER generate a sentiment lexicon through a phrase-level NLP tool~\cite{sigir14:efm,sigir18:mter}. Similarly, TriRank utilizes the extracted aspects to construct a tripartite graph over users and items~\cite{cikm15:trirank}. These works heavily rely on the performance of the external toolkit.

The second category aims to automatically infer explainable aspects from reviews by devising an internal component. JMARS learns the aspect representations of user and item by using topic modeling to conduct collaborative filtering~\cite{kdd14:jmars}. AFLM proposes an aspect-aware topic model over the reviews to learn diferent aspects of user and item in topic space~\cite{www18:alfm}. \eat{Then an aspect-aware latent factor model is devised to estimate the importance and rating of each aspect~\cite{www18:alfm}.} Very recently, ANR proposes a co-attention mechanism in the neural architecture to automatically estimate aspect-level ratings and aspect-level importance in an end-to-end fashion~\cite{cikm18:anr}.

Many of the above review-based and aspect-based solutions aim to identify the important words or aspects to enhance the recommendation and facilitate explanation. However, only an incomplete picture is drawn by them towards the complex reasoning process underlying a rating behavior. They are incapable of telling whether a user likes or dislikes an aspect of an item according to what viewpoint the user holds and to what extent. In this work, the proposed \baby can be seen as a preliminary step to fill in the missing part.

\subsection{Capsule Networks for NLP}
The capsule network is proposed as a hierarchical architecture to model the complex relations among latent features~\cite{icann11:capsule}. The affiliated dynamic routing (Routing by Agreement) mechanism in a capsule ensures that low-level features can be selectively aggregated to form high-level features~\cite{nips17:capsule}. Recently, this notion has been applied to some NLP tasks, including relation extraction~\cite{emnlp18:attCap}, text classification~\cite{emnlp18:yang}, zero-shot user intent detection~\cite{emnlp18:zeroshot} and multi-task learning~\cite{emnlp18:mcapsnet}. In this work, we exploit the capsule-based architecture with a new proposed Routing by Bi-Agreement (RBiA) mechanism to achieve multiple objectives jointly. RBiA calculates the inter-capsule and intra-capsule agreements to derive the output of a capsule, which would be a useful complement to the existing studies.

\section{The Proposed Model}\label{sec:alg}
\begin{figure*}
\centering
\includegraphics[height=56mm, width=.96\linewidth]{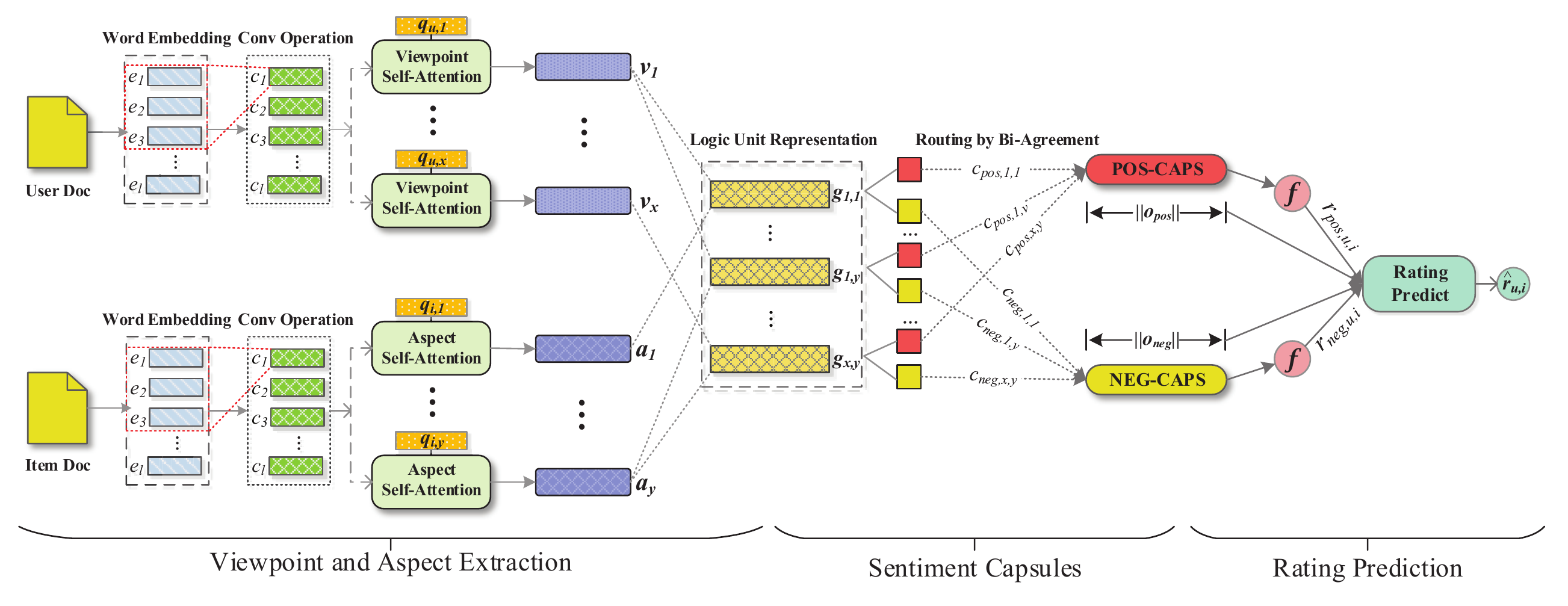}
\caption{The network architecture of \baby.}\label{fig:baby}
\end{figure*}

In this section, we present the proposed capsule network \baby and Figure~\ref{fig:baby} illustrates its overall architecture.

\subsection{Viewpoint and Aspect Extraction}\label{subsec:viewandaspect}
As previously mentioned, a user document $D_u$ is formed by merging all the reviews written by user $u$. And an item document $D_i$ can be formed in a similar way. We firstly \eat{utilize convolution operation, gating mechanism and self-attention to} extract the user veiwpoints and item aspects from the correponding documents respectively. Here, we give a detailed description about viewpoint extraction in \baby. The same procedure applies for aspect extraction. 

\paratitle{Context Encoding.} Given a user document $D_u=(w_1,w_2,...,w_l)$ where $l$ is the document length in number of words, we first project each word to its embedding representation: $\mathbf{D}_u=(\mathbf{e}_1,...,\mathbf{e}_l)$. Here, $\mathbf{e}_j\in R^d$ is the embedding vector for word at the $j$-th position, and $d$ is the embedding size. Then, to extract possible aspect-specific information around each word, we perform convolution operation with ReLU (\ie Rectified Linear Unit) as the activation function. Specifically, the context is encoded by using a window spanning $\frac{c - 1}{2}$ words on both side of each word. The resultant latent feature vectors are $[\mathbf{c}_1,...,\mathbf{c}_l]$ and $\mathbf{c}_j\in R^n$ is the latent contextual feature vector for $j$-th word, and $n$ is the number of filters.

\paratitle{Self-Attention.} Intuitively, not all words in the document are important for each viewpoint. To identify which features carried by $\mathbf{c}_j$ are relevant to each viewpoint, we leverage a viewpoint-specific gating mechanism:
\begin{align}
\mathbf{s}_{u,x,j} = \mathbf{c}_j\odot \sigma(\mathbf{W}_{x,1}\mathbf{c}_j + \mathbf{W}_{x,2}\mathbf{q}_{u,x} + \mathbf{b}_{x})\label{eqn:gate}
\end{align}
where $\mathbf{W}_{x,1},\mathbf{W}_{x,2}\in R^{n\times n}$ and $\mathbf{b}_{x}\in R^n$ are transform matrices and bias vector respectively for the $x$-th viewpoint, $\sigma$ is the sigmoid activation and $\odot$ is the element-wise product operation, $\mathbf{q}_{u,x}$ is the  embedding of $x$-th viewpoint shared for all users, which is learned by model optimization. We then extract the contextual viewpoint representation $\mathbf{p}_{u,x,j}$ from $\mathbf{s}_{u,x,j}$ through a projection: $\mathbf{p}_{u,x,j} = \mathbf{W}_p\mathbf{s}_{u,x,j} \label{eqn:transform}$
where $\mathbf{W}_p \in R^{k \times n}$ is a viewpoint-shared transform matrix. It is reasonable that the viewpoint-specific semantic of the same word within similar contexts could be different. For example, opposite sentiments are held by same word `long' in contexts ``the battery can sustains a \emph{long} time'' and ``the waiting time is \emph{long}''. The aspect-specific gating and projection help us extract the viewpoint relevant information precisely by disambiguating the semantics of each word.

It is natural that more viewpoint-specific words mentioned in the reviews, more firmly the viewpoint held by the user. For example, `performance', `durable', and `practical' would be frequently mentioned by a pragmatist while `lovely' and `aesthetics' are supposed to dominate the reviews written by appeareance-valuing users. In \baby, we propose to leverage the self-attention mechanism to derive the user viewpoint from her document. Specifically, We first derive the rudimentary representation of the viewpoint by taking the average sum of the constituent viewpoint-specific contextual embeddings in user document (\ie $\mathbf{v}_{u,x} = \frac{1}{l}\sum_{j}\mathbf{p}_{u,x,j}$). Then the intra-attention weights are calculated as $attn_{u,x,j} = softmax(\mathbf{p}_{u,x,j}^{\top}\mathbf{v}_{u,x})$. At last, the viewpoint $\mathbf{v}_{u,x}$ is represented as a weighted sum: 
\begin{align}
\mathbf{v}_{u,x} &= \sum_{j} attn_{u,x,j}\mathbf{p}_{u,x,j} \label{eqn:selfattn}
\end{align}
The intra-attention mechanism enables the viewpoint extraction to capture the features which are consistently important for different viewpoints. For model simplicity, we restrict the viewpoint number of a user and aspect number of an item to be the same. Following the same procedure, we extract $M$ aspects for an item from the corresponding item document with a different set of parameters (\eg the convolution operation, aspect embedding and gating mechanism).

\paratitle{Logic Unit Representation.} With the user viewpoints and item aspects extracted above, we need to identify the rules that a user applies when making a rating towards a specific item. Actually, for decision making, these rules take the form of various causes and their effects. It is reasonable that the user and the item should appear on both sides of a cause. Hence, a cause is formed by packing a user viewpoint with an item aspect together, which is named a \textit{logic unit}. Specifically, given the $x$-th viewpoint $\mathbf{v}_{u,x}$ of user $u$ and $y$-th aspect $\mathbf{a}_{i,y}$ of item $i$, the representation $\mathbf{g}_{x,y}$ of the corresponding logic unit $g_{x,y}$ is derived as follows:
\begin{align}
\mathbf{g}_{x,y}=[(\mathbf{v}_{u,x}-\mathbf{a}_{i,y}) \oplus (\mathbf{v}_{u,x}\odot\mathbf{a}_{i,y})]\label{eqn:logicunit}
\end{align} 
where $\oplus$ is the vector concatenation operation. Clearly, the second-order feature interactions adopted in Equation~\ref{eqn:logicunit} could offer more expressive power to encode the hidden relevance between a viewpoint and an aspect~\cite{sigir17:nfm,acl18:li}. We also opt for including the latent representations of $\mathbf{v}_{u,x}$ and $\mathbf{a}_{i,y}$ to represent $g_{x,y}$. In our empirical evaluation, however, no further performance gain is observed. 

\subsection{Sentiment Capsule}\label{subsec:sentiCaps}

Given there are $M$ viewpoints/aspects for each user/item, we can form $M^2$ logic units in total with random pairing for each user-item pair. However, not all possible logic units are plausible or make sense in real-world. Here, we define \textit{informative logic units} as the ones that are semantically plausible. That is, an informative logic unit should hold 
the explicit or implicit implication relation (or association) with respect to its constituent viewpoint and aspect in semantic sense. So one of our objective is to identify the informative logic units. Furthermore, we target at inferring whether a user likes and dislikes an item according to what informative logic units, and to what extent. Here, we propose a capsule architecture based on~\cite{nips17:capsule} to implement all these objectives in one go.

In detail, two sentiment capsules, namely positive capsule and negative capsule, are adopted to jointly choose some logic units as the informative ones and resolve their sentiments. Specifically, in each sentiment capsule, the latent sentiment features are derived for each logic unit as $\mathbf{t}_{s,x,y}=\mathbf{W}_{s,x,y}\mathbf{g}_{x,y}\label{eqn:tsxy}$, where $s\in\mathcal{S}$ and $\mathcal{S}=\{\text{pos},\text{neg}\}$ refers to the two sentiments, $\mathbf{W}_{s,x,y}\in R^{k\times 2k}$ is the weight matrix, and $\mathbf{t}_{s,x,y}$ is the latent feature vector for logic unit $g_{x,y}$. Then, the capsule with sentiment $s$ takes all feature vectors $\{\mathbf{t}_{s,x,y}|x,y\in 1...M\}$ as input and derive its output vector via an iterative dynamic routing process. Actually, we can consider the output of each sentiment capsule as a weighted sum over these feature vectors:
\begin{align}
\mathbf{s}_{s,u,i}=\sum_{x,y}c_{s,x,y}\mathbf{t}_{s,x,y}\label{eqn:ssui}
\end{align}
where $c_{s,x,y}$ is the coupling coefficient indicating the importance of logic unit $g_{x,y}$ with respect to determining sentiment $s$ (\ie like or dislike) and to what extent (ref. Equation~\ref{eqn:rsui}). That is, a sentiment capsule is expected to capture a user's overall opinion in that sentiment. Besides, a capsule also encodes the probability of the existence (or activation) of the concept represented by the capsule in terms of the length of its output vector~\cite{nips17:capsule}. Specifically, a nonlinear squashing function transforms $\mathbf{s}_{s,u,i}$ into $\mathbf{o}_{s,u,i}$ whose length falls in the range of $(0,1)$.
\begin{align}
\mathbf{o}_{s,u,i}=\frac{\|\mathbf{s}_{s,u,i}\|^2}{1+\|\mathbf{s}_{s,u,i}\|^2}\frac{\mathbf{s}_{s,u,i}}{\|\mathbf{s}_{s,u,i}\|}\label{eqn:squash}
\end{align}
where $\|\cdot\|$ denotes the length of a vector. Note that, in Equation~\ref{eqn:squash}, the orientation of $\mathbf{s}_{s,u,i}$ is retained in $\mathbf{o}_{s,u,i}$. We expect that $\mathbf{o}_{s,u,i}$ encodes the answer about to what extent user $u$ likes or dislikes item $i$. Also, we utilize the length of $\mathbf{o}_{s,u,i}$ to answer whether user $u$ holds sentiment $s$ against item $i$ in a probability perspective (\ie answer about whether user $u$ likes or dislikes item $i$).

\paratitle{Limitation of Routing by Agreement.} The value of $c_{s,x,y}$ in Equation~\ref{eqn:ssui} is calculated via an iterative Routing by Agreement algorithm. As used in original capsule architecture~\cite{nips17:capsule}, the core step is to iteratively calculate an agreement score $b_{s,x,y}$ indicating the relatedness between $\mathbf{t}_{s,x,y}$ and other feature vectors inside the same sentiment capsule. Then, a softmax operation is utilized to derive a logic unit's coupling coefficients with respect to the two capsules as follows:
\begin{align}
c_{s,x,y}=\exp(b_{s,x,y})/\sum_{s\in\mathcal{S}}\exp(b_{s,x,y})\label{eqn:csxy}
\end{align}
In Equation~\ref{eqn:csxy}, we can see that a logic unit will contribute itself more to the ouput of a sentiment capsule when it has relative more agreement with this capsule than the other one. Recall that we need to identify the informative logic units that represent the causes underlying a rating behavior. Based on Equation~\ref{eqn:csxy}, a non-informative logic unit would still pose a large blur to hurt the sentiment abstraction ability of the capsules. For example, a non-informative logic unit would produce negative agreements ($-0.05$ vs. $-0.9$) with the two capsules at the begining of the dynamic routing process. However, with the softmax operation in Equation~\ref{eqn:csxy}, the coupling coefficients are $0.7$ vs. $0.3$ respectively for this logic unit. The larger weight inevitably makes a lot of noise towards the first sentiment capsule, which causes a large adverse influence in later iterations (ref. Algorithm~\ref{alg:routing}). In this sense, \baby fails to uncover the reasoning process as well as to guarantee semantic explanation. For this reason, we introduce Routing by Bi-Agreement.

\paratitle{Routing by Bi-Agreement.} Let $\neg s$ denote the opposite sentiment compared to sentiment $s$, an appropriate dynamic routing process for our task should hold the following three properties: \textbf{(1)} Given $b_{s,x,y}$ is relative larger inside the capsule and $b_{s,x,y}>b_{\neg s,x,y}$, the value of $c_{s,x,y}$ should be larger as well; \textbf{(2)} Given $b_{s,x,y}$ is relative larger inside the capsule but $b_{s,x,y}<b_{\neg s,x,y}$, the value of $c_{s,x,y}$ should be small; \textbf{(3)} Given $b_{s,x,y}$ is relative smaller inside the capsule, the value of $c_{s,x,y}$ should be smaller also. Note that, given a user-item pair, the informative logic unit extracted from the related reviews would have different importance to explain a rating behavior. Also, each informative logic unit should be exclusively associated with a single sentiment. Here, we let \baby automatically identify the important informative logic units for a given user-item pair towards each sentiment, which is ruled by the first and second property in a review-driven manner. On the other side, non-informative logic units are expected to have very low agreement with both sentiment capsules, just as the example mentioned above. The third property could help us suppress their impact. 

\begin{algorithm}[t]
\caption{\textit{Routing by Bi-Agreement} Algorithm}
\label{alg:routing}
\SetKwInOut{Input}{input}
\SetKwInOut{Output}{output}
\Input{Iteration number $\tau$, feature vectors $\mathcal{T}=\{\mathbf{t}_{s,x,y}|s\in\mathcal{S},x\in 1...M,y\in 1...M\}$, $\mathcal{S}=\{\text{pos},\text{neg}\}$
}
\Output{Coupling coefficients $\{c_{s,x,y}\}$
}
\vspace{1ex}
\ForEach{$\mathbf{t}_{s,x,y}\in\mathcal{T}$}{
    $b_{s,x,y}=0$ \tcc*{Initialization}
}
\ForEach{iteration}{
    \ForEach{$\mathbf{t}_{s,x,y}\in\mathcal{T}$}{
       \tcc*[h]{A softmax over two sentiment capsules}\;
       $\check{c}_{s,x,y}=\exp(b_{s,x,y})/\sum_{s\in\mathcal{S}}\exp(b_{s,x,y})$\;
       \tcc*[h]{A softmax inside a sentiment capsule}\; 
       $\hat{c}_{s,x,y}=\exp(b_{s,x,y})/\sum_{j,k}\exp(b_{s,j,k})$\;
       \tcc*[h]{$L1$ normalization}\;
       $c_{s,x,y}=\sqrt{\check{c}_{s,x,y}\hat{c}_{s,x,y}}/\sum_{j,k}\sqrt{\check{c}_{s,j,k}\hat{c}_{s,j,k}}$\;
    }
    \ForEach{$s\in\mathcal{S}$}{
       $\mathbf{s}_{s,u,i}=\sum_{x,y}c_{s,x,y}\mathbf{t}_{s,x,y}$\;
       $\mathbf{o}_{s,u,i}=\text{squash}(\mathbf{s}_{s,u,i})$\tcc*{See Equation~\ref{eqn:squash}}
    }
    \ForEach{$\mathbf{t}_{s,x,y}\in\mathcal{T}$}{
       \tcc*[h]{Update the agreement}\;
       $b_{s,x,y}=b_{s,x,y}+\mathbf{t}_{s,x,y}^{\top}\mathbf{o}_{s,u,i}$\; 
    }
}
return $\{c_{s,x,y}\}$
\end{algorithm}

We propose Routing by Bi-Agreement (RBiA), a new iterative dynamic routing process that holds the above three properties. RBiA is devised to derive coupling coefficient $c_{s,x,y}$ by considering the relative magnitudes of $b_{s,x,y}$ in both inter-capsule and intra-capsule views together. The detail of this dynamic process is described in Algorithm~\ref{alg:routing}. At first, RBiA assigns the same initial agreement for each logic unit over the two sentiment capsules (Lines 1-2). Then, we calculate a candidate coupling coefficient $\check{c}_{s,x,y}$ based on the inter-capsule comparison (Line~5). On the other hand, another candidate coupling coefficient $\hat{c}_{s,x,y}$ is calculated similarly, but by comparing $b_{s,x,y}$ in an intra-capsule view (Line~6). We calculate coupling cofficient $c_{s,x,y}$ by performing $L1$ normalization over the geometric mean of $\check{c}_{s,x,y}$ and $\hat{c}_{s,x,y}$ inside each sentiment capsule (Line~7). The geometric mean is an appealing choice for our task since a percentage change in either $\check{c}_{s,x,y}$ or $\hat{c}_{s,x,y}$ will result in the same effect on the geometric mean. By considering the two different views on agreement $b_{s,x,y}$, it is clear to see that RBiA can satisfy the three properties mentioned above. Note that each coupling coefficient $c_{s,x,y}$ is updated based on the measured agreement between feature vector $\mathbf{t}_{s,x,y}$ and the output $\mathbf{o}_{s,u,i}$ at the current iteration (Line~12). The output vector $\mathbf{o}_{s,u,i}$ calculated at the last iteration is then used for rating prediction, which will be described in the following.

\subsection{Rating Prediction and Optimization}\label{ssec:optimization}

Given the output $\mathbf{o}_{s,u,i}$, we calculate rating $r_{s,u,i}$ user $u$ would give for item $i$ with sentiment $s$ as follows:
\begin{align}
r_{s,u,i} &= \mathbf{w}_s^{\top}\mathbf{h}_{s,u,i}+b_{s,3}\label{eqn:rsui}\\
\eta_s &= \sigma(\mathbf{H}_{s,1}\mathbf{o}_{s,u,i}+\mathbf{b}_{s,1})\label{eqn:eta}\\
\mathbf{h}_{s,u,i} &= \mathbf{\eta}_{s}\odot\mathbf{o}_{s,u,i}+(\mathbf{1}-\eta_{s})\odot tanh(\mathbf{H}_{s,2}\mathbf{o}_{s,u,i}+\mathbf{b}_{s,2})\label{eqn:hsui}
\end{align}
where $\mathbf{w}_s$ is the regression weight vector, $b_{s,3}$ is a bias term, $\mathbf{H}_{s,1}$, $\mathbf{H}_{s,2}\in R^{k\times k}$ are transform matrices, $\mathbf{b}_{s,1},\mathbf{b}_{s,2}\in R^{k}$ are the bias vectors, and $\mathbf{1}$ is the vector with all elements being $1$. By further considering the user and item biases, the overall rating score is then calculated as follows:
\begin{align}
\hat{r}_{u,i}=f_C(r_{pos,u,i}\cdot\|\mathbf{o}_{pos,u,i}\|-r_{neg,u,i}\cdot\|\mathbf{o}_{neg,u,i}\|)+b_u+b_i\label{eqn:rating}
\end{align}
where $\hat{r}_{u,i}$ is the predicted rating score, $b_u$ and $b_i$ are the corresponding bias for user $u$ and item $i$ respectively, and function $f_C(x)=1+\frac{C-1}{1+\exp(-x)}$ is a variant of the sigmoid function, producing a value within the target rating range of $[1,C]$. Here we explicitly consider the presence of sentiment $s$ for the given user-item pair in terms of vector length. Also, the utilization of one-layer highway network (ref. Equation~\ref{eqn:eta} and~\ref{eqn:hsui}) offers more flexibility to model complicated scenarios. For example, a user would dislike an item to some extent according to some cause . However, a high rating score would be also given by this user due to some more important causes (ref. Figure~\ref{fig:intro}).

\paratitle{Multi-task Learning.} For model optimization, we use Mean Square Error (MSE) as the objective to guide the parameter learning:
\begin{align}
L_{sqr}=\frac{1}{|\mathcal{O}|}\sum_{(u,i)\in \mathcal{O}}(r_{u,i}-\hat{r}_{u,i})^2\label{eqn:lsqr}
\end{align}
where $\mathcal{O}$ denotes the set of observed user-item rating pairs, $r_{u,i}$ is the observed rating score assigned by user $u$ to item $i$. Note that we explicitly include a particular sentiment analysis task in \baby. However, by matching the observed rating scores alone, the training of \baby would easily strap into a local optimum, which may not correctly reflect the fine-grained sentiments. Therefore, we introduce a sub-task to enhance the capacity of sentiment analysis in \baby. Specifically, a lower rating score suggests that the user certainly dislikes the item. Similarly, a higher score is rated by a user because of her most happiness on the item. Hence, we assign a label $s_{u,i}$ for each observed user-item rating pair. When rating $r_{u,i}>\pi$, $s_{u,i}=pos$; otherwise, $s_{u,i}=neg$. That is, we can split $\mathcal{O}$ into two disjoint subsets $\mathcal{O}_{pos}$ and $\mathcal{O}_{neg}$. And $\pi$ is a predefined threshold. We then utilize a sentiment analysis task as another objective function:
\begin{align}
L_{stm}=\frac{1}{|\mathcal{O}|}(\sum_{(u,i)\in \mathcal{O}}\max(0,\epsilon-\|\mathbf{o}_{s_{u,i},u,i}\|))\label{eqn:lstm}
\end{align}
where $\epsilon$ is a threshold indicating the probability that a user-item pair should hold for the corresponding sentiment. While Equation~\ref{eqn:lstm} guides the proposed \baby to learn the capacity of performing sentiment analysis more precisely, however, we need to consider the data imbalance issue. That is, the number of the user-item pairs with a lower rating score is significantly less than the couterpart (\ie $|\mathcal{O}_{pos}|\gg|\mathcal{O}_{neg}|$). This imbalance problem hurts the model from identifying the negative sentiment. To recruite more supervision regarding the negative sentiment to alleviate this problem, we exploit a mutual exclusion principle as follows:
\begin{align}
L_{stm} =& \frac{1}{|\mathcal{O}|}(\sum_{(u,i)\in \mathcal{O}}\max(0,\epsilon-\|\mathbf{o}_{s_{u,i},u,i}\|)\nonumber\\
+& \max(0,\|\mathbf{o}_{\neg s_{u,i},u,i}\|-1+\epsilon))\label{eqn:finallstm}
\end{align}
In Equation~\ref{eqn:finallstm}, each user-item pair should also meet the condition $\|\mathbf{o}_{\neg s_{u,i},u,i}\|\leq 1-\epsilon$. We know that this constraint is too strict to be true in reality. But more weak supervision about the negative sentiment is exploited for model optimization. In other words, the feedbacks regarding both the positive and negative sentiments are provided by each user-item pair. In our experiments, we observe that plausible results are gained by \baby with this principle during the prediction phase (ref. Section~\ref{ssec:analysis}). In this work, we set $\epsilon=0.8$. Considering both two objectives, the final objective for \baby is a linear fusion of $L_{sqr}$ and $L_{stm}$:
\begin{align}
L=\lambda\cdot L_{sqr}+(1-\lambda)\cdot L_{stm}\label{eqn:totalloss}
\end{align}
where $\lambda$ is a tunable parameter controling the importance of each subtask. We use RMSprop~\cite{lecture12rmsp} for parameter update in an end-to-end fashion.

\section{Experiments}\label{sec:exo}

\begin{table*}
\centering
\caption{Statistics of the seven datasets}
\setlength{\tabcolsep}{1.3mm}{
  \begin{tabular}{ccccccccc}
  \toprule
  Datasets & \# users & \# items & \# ratings & \# words per review & \# words per user & \# words per item & pos/neg ratio & density\\
  \midrule
  Musical Instruments & $1,429$ & $900$ & $10,261$ & $32.45$ & $141.32$ & $200.12$ & $7.28$ & $0.798\%$\\
 Office Products & $4,905$ & $2,420$ & $53,228$ & $48.15$ & $197.93$ & $229.52$ & $5.73$ & $0.448\%$ \\
  Digital Music & $5,540$ & $3,568$ & $64,666$ & $69.57$ & $216.21$ & $266.51$ & $4.14$ & $0.327\%$\\
  Tools Improvement & $16,638$ & $10,217$ & $134,345$ & $38.75$ & $162.53$ & $212.48$ & $5.42$ & $0.079\%$\\
  Video Games & $24,303$ & $10,672$ & $231,577$ & $72.13$ & $188.79$ & $260.60$ & $3.08$ & $0.089\%$\\
  Beer & $7,725$ & $21,976$ & $66,625$ & $17.31$ & $34.06$ & $103.20$ & $1.47$ & $0.039\%$\\
  Yelp16-17 & $167,106$ & $100,229$ & $1,217,208$ & $38.86$ & $133.60$ & $155.18$ & $3.29$ & $0.007\%$ \\
  \bottomrule \label{tab:stats}
  \end{tabular}}
\end{table*}

In this section, comprehensive experiments are conducted on seven datasets from three different sources to evaluate the performance of \baby\footnote{Our implementation is available at~\url{https://github.com/WHUIR/CARP}.}.

\subsection{Experimental Setup}\label{ssec:setup}
\paratitle{Datasets.}
In our experiment, we use seven datasets from three sources: Amazon-$5$cores\footnote{http://jmcauley.ucsd.edu/data/amazon/}~\cite{www16:dataset1}, Yelp, and the RateBeer website. For the Amazon-$5$cores dataset, five datasets from different domains are used (\ie \textit{Musical Instruments}, \textit{Office Products}, \textit{Digital Music}, \textit{Tools Improvement}, and \textit{Video Games}). The others are from the Yelp challenge website\footnote{https://www.yelp.com/dataset/challenge} (Round 11) and the RateBeer website\footnote{https://www.ratebeer.com/} (called \textit{Beer}), respectively. Note that, we also adopt $5$-core settings over these two datasets: each user and item has at least $5$ reviews. Besides, we only retain the records in Yelp spanning from $2016$ to $2017$ as the final dataset, denoted as \textit{Yelp16-17}. The target rating ranges used in these datasets are $[1,5]$\footnote{In Beer, we convert the rating range of the Overall Rating which is $[4, 20]$ into $[1, 5]$.}, we therefore set $C=5$ and $\pi=3$.

On all datasets, we adopt the same preprocessing steps used in~\cite{recsys16:convMf}. Then, we filter out the rating records which contain empty review afterwards. Detailed statistics of the seven preprocessed datasets are given in Table~\ref{tab:stats}. We can see that ratio $|\mathcal{O}_{pos}|/|\mathcal{O}_{neg}|$ (``pos/neg ratio'' in Table~\ref{tab:stats}) is very large on most datasets. For each dataset, we randomly build the training set and testing set in the ratio $80:20$. Moreover, $10\%$ records in the training set are randomly selected as the validation set for hyper-parameter selection. Note that at least one interaction for each user/item is included in the training set. The target reviews in the validation and testing sets \textit{are excluded} since they are unavailable in the practical scenario.

\begin{table*}
\small
\centering
\caption{Overall performance comparison on seven datasets in terms of \textit{MSE}. The best and second best results are highlighted in boldface and underlined respectively. $\blacktriangle\%$ denotes the relative improvement of \baby over the review-based alternatives. $\dagger$ indicates that the difference to the best result is statistically significant at $0.05$ level.}
\begin{tabular}{c||c|c|c|c|c|c|c}
\toprule
Method & Musical Instruments & Office Products & Digital Music & Tools Improvement &  Video Games & Beer & Yelp$16$-$17$\\
\midrule
PMF & $1.398^\dagger$ & $1.092^\dagger$ & $1.206^\dagger$ & $1.566^\dagger$ & $1.672^\dagger$ & $1.641^\dagger$ & $2.574^\dagger$ \\
\hline\hline
RBLT & $0.815^\dagger$ & $0.759^\dagger$ & $0.870^\dagger$ & $0.983^\dagger$ & $1.143^\dagger$ &  $\underline{0.576}^\dagger$ & $1.569^\dagger$\\
\hline
CMLE  & $0.817^\dagger$ & $0.759^\dagger$ & $0.885^\dagger$ & $1.020^\dagger$ & $1.253^\dagger$  & $0.605^\dagger$ & $1.593^\dagger$\\
\hline
DeepCoNN & $0.814^\dagger$ & $0.860^\dagger$ & $1.056^\dagger$ & $1.061^\dagger$ & $1.238^\dagger$ &  $0.618^\dagger$ & $1.593^\dagger$ \\
\hline
D-Attn & $0.982^\dagger$ & $0.825^\dagger$ & $0.911^\dagger$ & $1.043\dagger$ & $\underline{1.145}^\dagger$  & $0.614^\dagger$ & $1.573^\dagger$ \\
\hline
TransNet & $0.798^\dagger$ & $0.759^\dagger$ & $0.913^\dagger$ & $1.003^\dagger$ & $1.190^\dagger$ & $0.587^\dagger$  & $\underline{1.523}^\dagger$\\
\hline
TARMF & $0.943^\dagger$  & $0.789^\dagger$ & $\underline{0.853}^\dagger$ & $1.169^\dagger$ & $1.195^\dagger$ & $0.912^\dagger$ & $1.914^\dagger$ \\
\hline
MPCN & $0.824^\dagger$ & $0.769^\dagger$ & $0.903^\dagger$ & $1.017^\dagger$ & $1.201^\dagger$ & $0.616^\dagger$ & $1.617^\dagger$ \\
\hline
ANR & $\underline{0.795}^\dagger$ & $\underline{0.742}^\dagger$ & $0.867^\dagger$ & $\underline{0.975}^\dagger$ & $1.182^\dagger$ & $0.590^\dagger$ & $1.553^\dagger$ \\
\midrule
\baby & $\textbf{0.773}$ & $\textbf{0.719}$ & $\textbf{0.820}$ & $\textbf{0.960}$ & $\textbf{1.084}$ & $\textbf{0.556}$ & $\textbf{1.508}$\\
\hline
$\blacktriangle\%$ & $2.8-21.3$ & $3.1-16.4$ & $3.9-22.3$ & $1.5-17.9$ & $5.3-13.5$ & $3.5-39.0$ & $0.98-21.2$\\
\hline\hline
\baby-RA & $0.789$ & $0.727$ & $0.836$ & $0.969$ & $1.100$ & $0.567$ & $1.536$\\
\bottomrule
\end{tabular} \label{tab:comparison}
\end{table*}

\begin{table*}
\small
\centering
\caption{Impact of different viewpoint/aspect numbers in \baby. The best results are highlighted in boldface.}
\begin{tabular}{c||c|c|c|c|c|c|c}
\toprule
$M$ & Musical Instruments & Office Products & Digital Music & Tools Improvement & Video Games & Beer & Yelp$16$-$17$\\
\midrule
$3$ & $0.776$ & $0.728$ & $0.824$ & $0.961$ & $1.090$ & $0.560$ & $1.513$ \\
$5$ & $0.776$ & $0.716$ & $\textbf{0.818}$ & $0.962$ & $1.087$ & $\textbf{0.552}$ & $\textbf{1.503}$ \\
$7$ & $\textbf{0.771}$ & $\textbf{0.715}$ & $\textbf{0.818}$ & $0.958$ & $1.083$ & $0.564$ & $1.517$ \\
$9$ & $0.778$ & $0.726$ & $0.819$ & $\textbf{0.950}$ & $\textbf{1.074}$ & $0.562$ & $1.509$\\
\bottomrule
\end{tabular} 
\label{tbl:Manalysis}
\end{table*} 

\paratitle{Baselines.} Here, we compare the proposed \baby against the conventional baseline and recently proposed state-of-the-art rating prediction methods: (a) probabilistic matrix factorization that leverages only rating scores, \textbf{PMF}~\cite{nips07:pmf}; (b) latent topic and shallow embedding learning models with reviews, \textbf{RBLT}~\cite{ijcai16:rblt} and \textbf{CMLE}~\cite{ijcai16:cmle}; (c) deep learning based solutions with reviews, \textbf{DeepCoNN}~\cite{wsdm17:deepconn}, \textbf{D-Attn}~\cite{recsys17:dattn}, \textbf{TransNet}~\cite{recsys17:transnet}, \textbf{TARMF}~\cite{www18:tarmf}, \textbf{MPCN}~\cite{kdd18:mpcn} and \textbf{ANR}~\cite{cikm18:anr}. Among these methods, D-Attn, TARMF, MPCN and ANR all identify important words for rating prediction. Note that there are many other state-of-the-art models, such as HFT~\cite{recsys13:hfht}, EFM~\cite{sigir14:efm}, JMARS~\cite{kdd14:jmars}, CDL~\cite{kdd15:cdl}, ConvMF~\cite{recsys16:convMf} and ALFM~\cite{www18:alfm}. These works have been outperfomed by one or several baselines compared here. Hence, we omit further comparison for space saving.

\paratitle{Hyper-parameter Settings.}
We apply grid search to tune the hyper-parameters for all the methods based on the setting strategies reported in their papers. The final performances of all methods are reported over 5 runs. The latent dimension size is optimized from $\{25,50,100,150,200,300\}$. The word embeddings are learned from scratch. The dimension size of word embedding is set to $300$ (\ie $d=300$). The batch size for Musical Instruments, Beer, Office Products and Digital Music is set to $100$. For other bigger datasets, the batch size is set to $200$. The number of convolution filters is set to $50$ for convolution based methods (including \baby, $n=50$).

For \baby, window size is $c=3$, the iteration number $\tau$ is set to $3$ for Routing by Bi-Agreement, the number of viewpoints or aspects for each user/item is set to be $5$ (\ie $M=5$), and $\lambda$ is $0.5$. We set the keep probability of dropout to be $0.9$ and learning rate to $0.001$ for model training. The dimension size $k$ is set to be $25$\footnote{We observe that $k$ is optimal in $[25,100]$ on all datasets.}.

\paratitle{Evaluation Metric.}
Here, we use \textit{MSE} (ref. Equation~\ref{eqn:lsqr}) as performance metric, which is widely adopted in many related works for performance evaluation~\cite{www18:tarmf,cikm18:wu,kdd18:mpcn}. The statistical significance test is conducted by performing the student \textit{t-test}.

\subsection{Performance Evaluation}\label{ssec:evaluation}
A summary of results of all methods over the seven datasets are reported in Table~\ref{tab:comparison}. Several observations can be made. First, it is not surprising that the interaction-based method (PMF) consistently yields the worst on all seven datasets. This observation is consistent with what have been made in many review-based works~\cite{recsys16:convMf,wsdm17:deepconn,recsys17:dattn}. 

Second, among review-based baselines, there is no dominating winner acorss seven datasets. \eat{ConvMF and }DeepCoNN consistently performs worse than the other solutions across different datasets. This is reasonable since neither word-level nor aspect-level attention mechanism is used for the feature extraction. In contrast, a much better performance is obtained by D-Attn against DeepCoNN in most datasets, due to the dual word-level attention mechanisms utilized in the former. Also, TransNet performs significantly better than DeepCoNN in all the datasets. This observation is consistent with prior works~\cite{recsys17:transnet,kdd18:mpcn}. By repeating each review according to its rating score (\ie rating boosting), RBLT can precisely extract the semantic topics from the textual information. Relative good performance (\ie comparable to the best baseline) is achieved across different datasets, especially on the sparser ones. ANR performs the best in three datasets while TransNet, D-Attn, TARMF and RBLT outperform the others in each of the rest four datasets. Note that ANR models semantic information from reviews in the aspect-level. This suggests that the fine-grained representation learning is a promising avenue towards better understanding of a rating behavior.

Third, as Table~\ref{tab:comparison} shows, \baby consistently achieves the best performance across the seven datasets. It is worthwhile to highlight that a significant improvement is gained by \baby on Beer which is the second sparsest dataset with least review information (ref. Table~\ref{tab:stats}). Compared with three recently proposed state-of-the-art models (\ie TARMF, MPCN and ANR), \baby obtains up to $39.0\%$, $9.7\%$ and $8.0\%$ relative improvement respectively. Overall, the experimental results demonstrate that \baby is effective in modeling a rating behavior from reviews for rating prediction.

\begin{figure}[t]
\centering
\begin{subfigure}[Ratio values on different ranks]{\includegraphics[width=.46\linewidth]{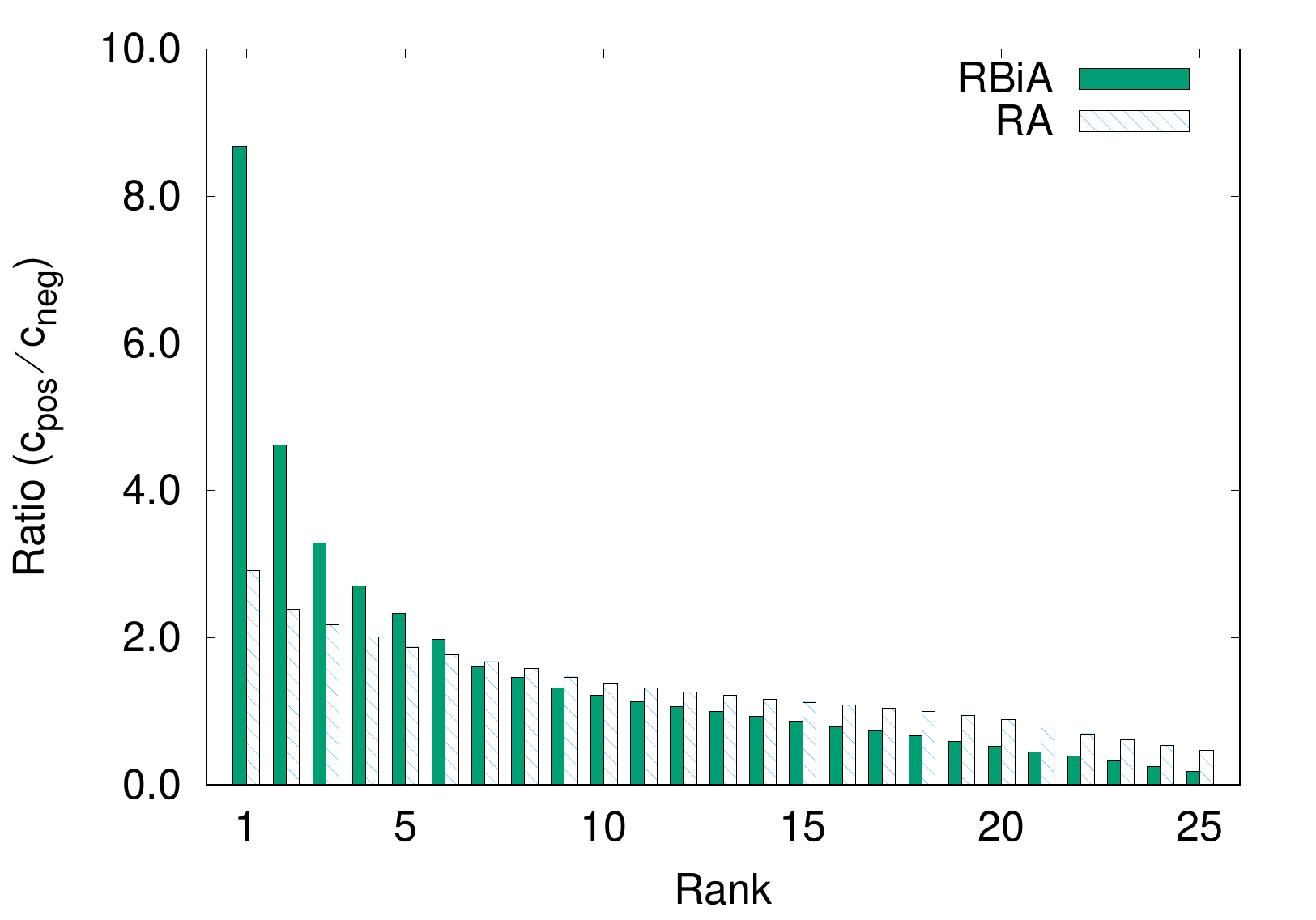}
   \label{fig:ratio}
 }%
\end{subfigure}\hfill
\begin{subfigure}[Performance on differnt $\lambda$ values]{\includegraphics[width=.46\linewidth] {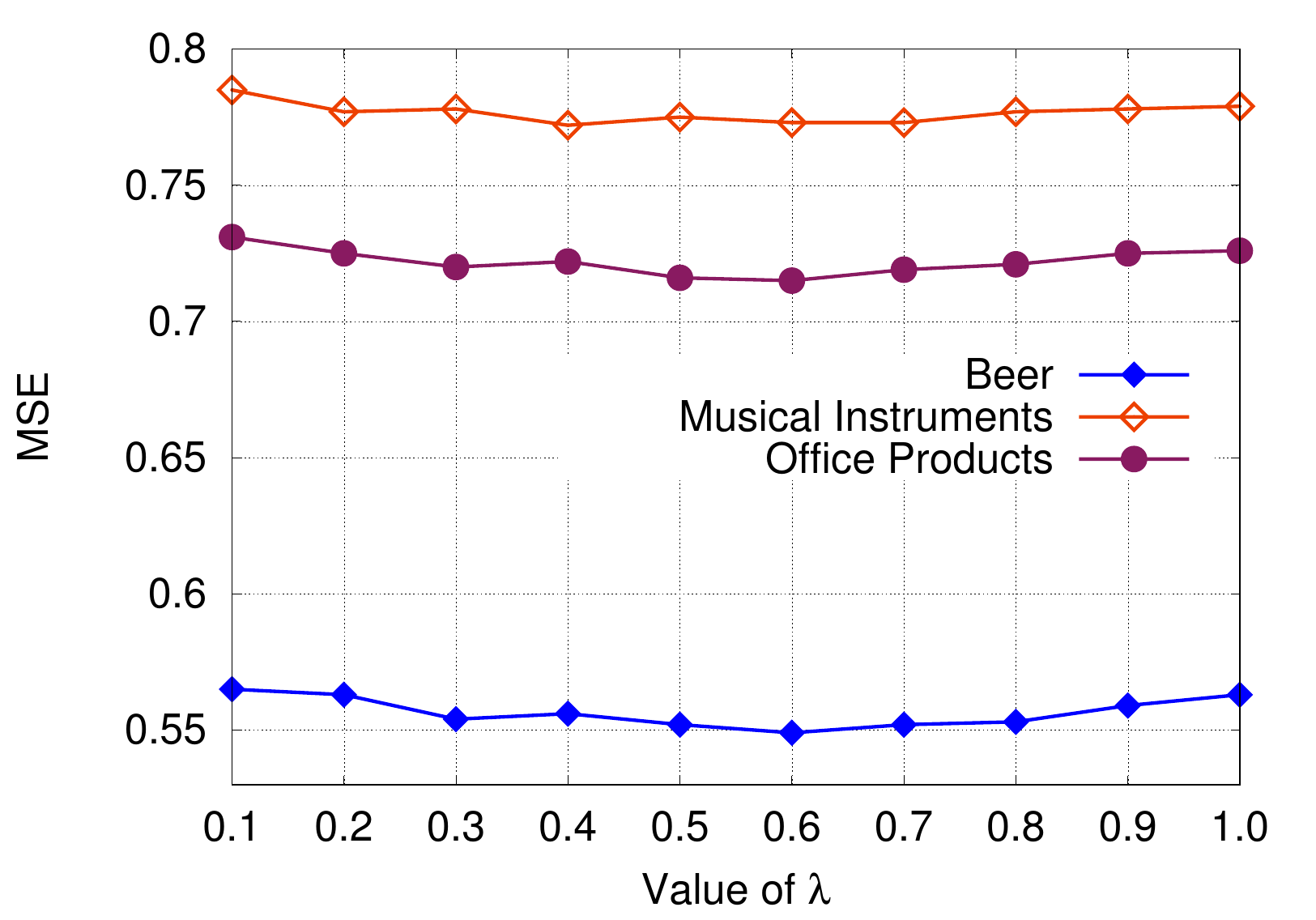}
   \label{fig:lamb}
 }%
\end{subfigure}
\caption{Ratios at different ranks (a), and performance with $\lambda$ values (b).}\label{fig:lambda}
\end{figure}

\subsection{Analysis of \baby}\label{ssec:analysis}
We further investigate the impact of the parameter settings (\ie $\tau,M,\lambda$) to the performance of \baby on the validation set. When studying a parameter, we fix the other parameters to the values described in Section~\ref{ssec:setup}. Note that not all performance patterns in all datasets are presented below. Since the similar pattern is observed and performance scales are quite different, we omit some of them for space saving. 

\begin{table}
\scriptsize
\centering
\caption{Impact of different numbers of iterations in RBiA. The best results are highlighted in boldface.}
\setlength{\tabcolsep}{1.2mm}{
\begin{tabular}{c||c|c|c|c|c}
\toprule
$\tau$ & Musical Instruments & Office Products & Digital Music & Tools Improvement & Beer\\
\midrule
$1$ & $0.786$ & $0.733$ & $0.852$ & $0.976$ & $0.573$ \\
$2$ & $0.778$ & $0.720$ & $0.833$ & $0.963$ & $0.557$ \\
$3$ & $0.776$ & $\textbf{0.716}$ & $\textbf{0.818}$ & $0.962$ & $\textbf{0.552}$ \\
$4$ & $\textbf{0.772}$ & $0.727$ & $0.827$ & $\textbf{0.954}$ & $0.555$ \\
\bottomrule
\end{tabular}} \label{tbl:itr}
\end{table}

\begin{table*}
\scriptsize
\centering
\caption{Example study of three user-item pairs from Office Products and Musical Instruments.}
\begin{tabular}{c||c|p{11.8cm}|c|c}
\toprule
\multicolumn{3}{l|}{$\text{user}_{1}$ - $\text{item}_{1}$ (Smead MO File Box): $r_{1,1}= 2.0,\hat{r}_{1,1} = 3.03,\|\mathbf{o}_{pos,1,1}\| = 0.589,\|\mathbf{o}_{neg,1,1}\| = 0.460,r_{pos,1,1} = 0.761,r_{neg,1,1} = 0.735$} & $c_{pos,x,y}$ & $c_{neg,x,y}$
\\ \hline

\multirow{2}{*}{$g_{2,2}$} & \textbf{viewpoint}: & \multicolumn{1}{l|}{The base is \hlgol{solid metal} with \hlgol{high quality casters}.} & \multirow{2}{*}{$0.034$} & \multirow{2}{*}{$\mathbf{0.079}$}
\\ 

& \textbf{aspect}: & \multicolumn{1}{l|}{One last thing, I think is \hlgol{holds light stuffs}, don`t \hlgol{over loading it}.}  & & 
\\ \hline

\multirow{2}{*}{$g_{1,5}$} & \textbf{viewpoint}: & \multicolumn{1}{l|}{These labels \hlgol{save me money} by letting me reuse old folders.} & \multirow{2}{*}{$0.026$} & \multirow{2}{*}{$\mathbf{0.064}$}
\\ 

& \textbf{aspect}: & \multicolumn{1}{l|}{My \hlgol{concerns is the price}. I think is \hlgol{seriously overprice} for a cardboard box. $\cdots$} & & 
\\ \hline

\multirow{2}{*}{$g_{5,3}$} & \textbf{viewpoint}: & \multicolumn{1}{l|}{These are \hlgol{great organizational tools}, whether to \hlgol{replace lost tabs} or to \hlgol{relabel a whole drawer} in your own unique manner.} & \multirow{2}{*}{$\mathbf{0.074}$} & \multirow{2}{*}{$0.029$}
\\ 

& \textbf{aspect}: & \multicolumn{1}{l|}{I could easily use a bunch of these to \hlgol{organize} \hlbl{papers and magazines} at work and at home} & & 
\\ \hline

\multicolumn{2}{c}{\textbf{target review}} & \multicolumn{3}{m{11.5cm}}{$\cdots$ The Smead MO File Box seemed like \ulred{it could be a good way to keep those papers right at hand.} $\cdots$ It's \ulgrn{a little disapointing for the price} that it doesn't come with a slip-cover. $\cdots$ The overall would not recommend it for general use \ulgrn{unless they reassess the price}.} 
\\ \midrule\midrule

\multicolumn{3}{l|}{$\text{user}_{1}$ - $\text{item}_{2}$ (Post-It Big Pad): $r_{1,2}= 4.0,\hat{r}_{1,2} = 4.45,\|\mathbf{o}_{pos,1,2}\| = 0.726,\|\mathbf{o}_{neg,1,2}\| = 0.293,r_{pos,1,2} = 0.737,r_{neg,1,2} = 0.651$} & $c_{pos,x,y}$ & $c_{neg,x,y}$
\\ \hline

\multirow{2}{*}{$g_{3,4}$} & \textbf{viewpoint}: & \multicolumn{1}{l|}{It is also \hlgol{visually attractive with a silver back support}, and would look great in any office.} & \multirow{2}{*}{$\mathbf{0.078}$} & \multirow{2}{*}{$0.027$}
\\ 

& \textbf{aspect}: & \multicolumn{1}{l|}{Positives: \hlgol{Vivid yellow and large footprint} and ability to stick to surfaces means your message will be received. } & & 
\\ \hline

\multirow{2}{*}{$g_{1,1}$}  & \textbf{viewpoint}: & \multicolumn{1}{m{11.5cm}|}{Each tab has \hlgol{adhesive that holds well to the page} but can also be removed if needed.} & \multirow{2}{*}{$\mathbf{0.071}$} & \multirow{2}{*}{$0.017$}
\\ 

& \textbf{aspect}: & \multicolumn{1}{m{11.5cm}|}{The \hlbl{adhesive} is \hlgol{wide enough to securely hold the paper} on many different surfaces} & & 
\\ \hline

\multirow{2}{*}{$g_{2,1}$} & \textbf{viewpoint}: & \multicolumn{1}{l|}{The base is \hlgol{solid metal} with \hlgol{high quality casters}.} & \multirow{2}{*}{$0.018$} & \multirow{2}{*}{$\mathbf{0.083}$} 
\\ 

& \textbf{aspect}: & \multicolumn{1}{l|}{I want to cover some of my notes I had \hlgol{pre-written on a white board}.} & &
\\ \hline

\multicolumn{2}{c}{\textbf{target review}} & \multicolumn{3}{m{11.5cm}}{$\cdots$ The stick-em doesn't appear any stickier than a standard post-it, so I suspect these \ulred{extra large strips are needed to hold it on the wall}. $\cdots$ These large pads will definitely serve a purpose in these sessions, \ulred{allowing larger print that is easier to read}.} 
\\ \midrule\midrule

\multicolumn{3}{l|}{$\text{user}_{2}$ - $\text{item}_{3}$ (Guitar Patch Cable): $r_{2,3}= 5.0,\hat{r}_{2,3} = 4.56,\|\mathbf{o}_{pos,2,3}\| = 0.759,\|\mathbf{o}_{neg,2,3}\| = 0.295,r_{pos,2,3} = 0.941,r_{neg,2,3} = 0.753$} & $c_{pos,x,y}$ & $c_{neg,x,y}$
\\ \hline

\multirow{2}{*}{$g_{1,4}$} & \textbf{viewpoint}: & \multicolumn{1}{m{11.5cm}|}{I found that these were OK but the \hlgol{heads were bigger than I could stand to use} on my \hlgol{cramped for space} pedal board. } & \multirow{2}{*}{$\mathbf{0.124}$} & \multirow{2}{*}{$0.0198$}
\\ 

& \textbf{aspect}: & \multicolumn{1}{m{11.5cm}|}{I use these patch cables on my pedal board and they \hlgol{save valuable space}} & & 
\\ \hline

\multirow{2}{*}{$g_{4,3}$} & \textbf{viewpoint}: & \multicolumn{1}{m{11.5cm}|}{The delay pedal was \hlgol{so cheap} that I could not \hlgol{pass it up}. And let me tell you, it is \hlgol{easy and fun to use}} & \multirow{2}{*}{$\mathbf{0.090}$} & \multirow{2}{*}{$0.035$}
\\ 

& \textbf{aspect}: & \multicolumn{1}{l|}{\hlgol{Good quality} in spite of \hlgol{their reasonable price}. } & & 
\\ \hline

\multirow{2}{*}{$g_{3,1}$} & \textbf{viewpoint}: & \multicolumn{1}{l|}{\hlbl{Love} \hlgol{the heavy cord} and \hlgol{gold connectors}. } & \multirow{2}{*}{$0.019$} & \multirow{2}{*}{$\mathbf{0.065}$}
\\ 

& \textbf{aspect}: & \multicolumn{1}{l|}{Plus, \hlgol{they look nice}, \hlgol{unlike the various colors} that \hlgol{come with the bulkier}, cheaper 1'cords} & & 
\\ \hline

\multicolumn{2}{c}{\textbf{target review}} & \multicolumn{3}{m{11.5cm}}{\ulred{compact heads} allow more pedals to be loaded onto your overpriced pedal board. \ulred{the quality is good}. $\cdots$} \\
\bottomrule
\end{tabular} \label{tbl:showcase}
\end{table*}

\paratitle{Impact of Dynamic Routing.} We investigate the impact of $\tau$ in the dynamic routing of \baby. Table~\ref{tbl:itr} reports the performance pattern of \baby with varying iteration numbers in terms of \textit{MSE}. It is clear that more than two iterations leads to better prediction accuracy. The optimal performance is obtained by performing either $3$ or $4$ iterations. Based on the result, we perform three iterations in \baby (\ie $\tau=3$). We also conduct the experiments with the standard Routing by Agreement (RA) as proposed in~\cite{nips17:capsule}. Last row of Table~\ref{tab:comparison} reports the performance over seven datasets with this setting (\ie \baby-RA). It is found that the performance deteriorates to different extent. For each logic unit $g_{x,y}$ ranked at $i$-th position in the positive capsule, we further check the corresponding ratio $c_{pos,x,y}/c_{neg,x,y}$. Figure~\ref{fig:ratio} reports the averaged ratio for each position in the positive capsule calculated by RBiA and RA respectively over the testing set of Musical Instruments. We can see that the coupling coefficients of logic units calculated by RBiA are more sharp. This suggests that RBiA is more appropriate to identify the informative logic units that are highly relevant to a user's rating. Similar patterns are also observed in the other datasets. Overall, the results suggest the superiority of the proposed RBiA by suppressing the adverse impact from the non-informative logic units.

\paratitle{Impact of $M$ Value.}
The $M$ value specifies the number of viewpoints/aspects extracted for each user/item respectively. Here, we report the performance patterns of \baby by tuning $M$ amongst $\{3,5,7,9\}$. As shown in Table~\ref{tbl:Manalysis}, the optimal $M$ value is not consistent across the datasets. It seems that $M=5/7$ is desired in more datasets. Given the performance variation is small, we choose to use $M=5$ in our experiments.

\paratitle{Impact of Multi-task Learning.} Recall in Equation~\ref{eqn:totalloss}, parameter $\lambda$ controls the tradeoff in the multi-task learning setting of \baby. Figure~\ref{fig:lamb} plots the performance patterns of three datasets by tuning $\lambda$ in the range of $[0.1,1]$ with a step of $0.1$. When $\lambda=0$, there is not supervision towards how to predict the final rating scores. We observe that the optimal performance is consistently achieved when $\lambda$ is around $0.5$. We also observe that the performance becomes increasingly worse when $\lambda\rightarrow 1$. This validates the positive benefit of the proposed multi-task learning process in \baby. Based on the results, we fix $\lambda$ to be $0.5$ in our experiments, though this may not be an optimal setting for some datasets. Note that we also introduce a mutual exclusion principle in the multi-task learning process (ref. Equation~\ref{eqn:lstm} and~\ref{eqn:finallstm}). Without this mutual exclusion, \baby produces a worse performance which is comparable with the setting of $\lambda=1$. \eat{Also, almost all testing user-item pairs have the same vector length for two sentiment capsules under this setting, which is irrational in real-world.}

\subsection{Explainability Analysis}\label{ssec:explain}
We further check whether \baby can discover proper causes and their effects to interpret a rating behavior. 

Table~\ref{tbl:showcase} displays these sentences and the auxiliary information provided by \baby. Recall that \baby has encoded contextual information to represent each word. To better visualize a viewpoint (an aspect), we retrieve the top-$K$ phrases whose weight is the sum of the weights (\ie $attn_{u,x,j}$ in Equation~\ref{eqn:selfattn}) of the constituent words in the convolutional window. Finally, we pick the sentences containing these informative phrases to represent the corresponding viewpoint/aspect. Here, we choose $K=30$. We randomly sample three user-item pairs from two datasets. The first two pairs with the same user but different items are from Office Products. The last one is picked from Musical Instruments. For a pair with a higher rating, we list top-$2$ and top-$1$ logic units extracted by positive and negative sentiment capsules respectively, and vice versa. Note that an extracted viewpoint (or aspect) could cover several sub-viewpoints (or sub-aspects), given our $M$ is a global parameter. Also, it is observed that some phrases would appear in two viewpoints or aspects. This is reasonable since a viewpoint or aspect is a high-level concept composited by different phrases, some of which have just general semantics. We choose one or two of the most informative sentences of each logic unit. The informative phrases are highlighted by orange color. The stop words inside a context is also highlighted for better understand. The blue color indicates that the phrase is associated with more than one viewpoint/aspect. We also transform raw $r_{s,u,i}$ value by Max-Normalization to ease interpretation, since the final rating is estimated through a nonlinear transfer (ref. Equation~\ref{eqn:rating}). As a reference, we display the parts matched well by these logic units in target user-item review with red and green underlines for positive and negative sentiment capsules respectively.

\textbf{Example 1}: the item in the first pair is a Smead MO File Box with a low rating (\ie $2.0$) from the user. The first logic unit suggests that the user prefers metal texture and high quality. However, the associated aspect shows that this item can not hold heavy things. The second logic unit directly indicates that the user is a saver and the item is obviously overprice. There is also some thing the user likes about this item. The third logic unit shows that the user loves the flexiblity and versatility, which is indeed expressed in item document. However, this is a weak positive effect. Note that the causes and effects expressed by logic units $g_{1,5}$ and $g_{5,3}$ are really mentioned in the target review.

\textbf{Example 2}: compared with the first pair, the second one illustrates the causes and effects towards a Post-it Big Pads by the same user with a high rating (\ie $4.0$). The first and second logic units suggest that some information needs (\ie attractive appearance and convenient to post something) are matched well by the item. The target review written by the user also covers these two logic units very well. The third logic unit is not straightforward to interpret. Some implication, however, indeed exists underlying this logic unit. As displayed by the second logic unit in the first example, this user mentions metal texture and high quality (the same viewpoint in two logic units), indicating she also likes solid and invulnerable things. Meanwhile, it is true that a Post-it note does not have the property. The short length of the output of the negative capsule (\ie $0.293$)  indicates that this cause has a weak negative effect, which is also confirmed by the overall high rating from the user.

\textbf{Example 3}: this example is from Musical Instruments. The first logic unit shows that the user-item pair reaches an agreement on space saver. Similarly, the second logic unit demonstrates that the user holds a cost-effective viewpoint which is well matched by the high cost performance of the item. This logic unit is also well matched in the target review. The third logic unit involves a bit reasoning. It is obvious that the user prefers something with cool appearance. In contrast, this item looks uninteresting based on the review. The high rating indicates that this logic unit is not important. This effect is also captured by \baby with a relative lower $r_{neg,u,i}$ (\ie $0.753$) and a small vector length (\ie $0.295$).

Overall, our results demonstrate that \baby is superior in rating prediction and facilitates explanation with causes and effects at a finer level of granularity.

\vspace{-0.3em}
\section{Conclusion}\label{sec:con}
The proposed capsule network based model for recommendation and explanation can be seen as a single step towards reasoning a particular kind of human activity. Our extensive experiments demonstrate that better recommendation performance and understanding are obtained by modeling the causes and effects in fine-grained level. Actually, reasoning a rating behavior in terms of user viewpoints, item aspects and sentiments could nourish further benefits in E-Commerce, including cross-domain recommendation, user profiling, and user-oriented review summarization. We will invesigate these possibilities in the future.

\begin{acks}
This research was supported by National Natural Science Foundation of China (No.~61872278, No.~91746206). Chenliang Li is the corresponding author.
\end{acks}

\bibliographystyle{ACM-Reference-Format}
\balance
\bibliography{acmart}

\end{document}